\documentclass[page-classic]{epl2arXiv} 
\usepackage{amssymb}
\usepackage{graphics}
\usepackage{subfig}
\usepackage{epsfig}
\usepackage{color}

\title{Do planetary seasons play a fundamental role in attaining habitable climates?}

\shorttitle{Seasons and habitable climates}

\author{Kasper W. Olsen\thanks{Email: \email{kasol@nanotech.dtu.dk}} \and 
Jakob Bohr
}
\shortauthor{K.W. Olsen and J. Bohr}

\institute{                    
DTU Nanotech,
Building 344, 
\O rsteds Plads,
Technical University of Denmark,
DK-2800 Kongens Lyngby, Denmark
}

\pacs{96.}{Solar systems; planetology}

\abstract{
A simple phenomenological account for planetary climate instabilities is presented.
The description is based on the standard model where the balance of incoming stellar radiation and outward thermal radiation is described by the effective planet temperature. Often, it is found to have three different points, or temperatures, where the influx of radiation is balanced with the out-flux, even with conserved boundary conditions. Two of these points are relatively long-term stable, namely the point corresponding to a frozen-climate and the point corresponding to a hot-climate. The hypothesis promoted in this paper is the possibility that it is the intermediate third point which is the basis for habitable-climates. I.e. that this initially unstable point is made relatively stable over a long period by the presence of seasonal climate variations. This points to the axial inclination, and perhaps the presence of orbital eccentricity, as the origin of the stability of the habitable point. 
An analysis involving the inclination, the size of the ice caps, and the length of the year shows that within the currently accepted value of the heat capacity of the Earth, the otherwise unstable habitable point is stabilized.}

\begin{document}

\maketitle

\section{Introduction}

Since the advent of the first modern climate models in the 1960's \cite{budyko1969,sellers1969}, the past climate of the Earth has been a subject of intense debate.
One possibility is that at a point in the past the Earth has been fully, or at least nearly fully, covered by ice \cite{kirschivink1992,hoffman1998}. This is the so-called Snowball Earth hypothesis.
The scientific debate continues as to whether this hypothesis is true or not \cite{kennedy1998,hoffman2002}. Regardless of the final outcome of this debate, it is worthwhile to notice that the debate would not have set off if it was not for the pioneering insight that {\it the balance between incoming heat and outgoing heat can have more than one solution}. This revelation has been an early stimulation for the work on current climate models of the Earth \cite{kaper2013}. See also data on radiative forcing and albedo feedback \cite{flanner2011}.
Recent work also considers the effects of daily cycles (diurnal asymmetry) \cite{davy2016}.

In this paper we discuss the nature of the planetary heat balance in its most simple and conceptual form. 
The results therefore apply to planet Earth, as well as to exoplanets in the habitable zone \cite{kasting1993}. 
We shall revisit the question of which of the possible solutions to the heat balance equation
correspond to a habitable climate. In the case of three such points we will designate them the planetary snowball point, the habitable point, and the desert planet point by increasing temperature. Hence, what we suggest is a reinterpretation, the suggestion is that what is normally considered the unstable point is actually the habitable point. This assignment follows from the "albedo curve" taking habitable to be in the coexistence regime of water and ice. 
At first, it may seem that only the snowball point and the desert point are stable equilibrium points \cite{kaper2013}. This interpretation is based on a static analysis of how the two power terms develop when one moves to a point slightly away from the equilibrium points. In general, the way one consider dynamical systems was pioneered by Lyapunov \cite{lyapunov1882} and has later led to a significant impact on the understanding of dynamical systems, attractors, and Lyapunov exponents \cite{thompson2001,cvitanovic2012}. One crucial observation is that under certain conditions forced oscillations can stabilize an otherwise unstable equilibrium, e.g. as in the dynamically driven hill-top stability \cite{zakrzhevsky2011}.

We suggest that the habitable point can locally become stabilized for a long extended period of time by the climate fluctuations which arise from seasonal changes that are caused by the axial inclination of the planet. I.e. that seasons play an important role for climate stability. A simplified stability analysis suggest that a significant axial inclination is needed.\\

\section{Method}

Following the standard model \cite{kaper2013}, a heuristic model that demonstrates the kind of stability one can obtain for the heat balance can be developed. The planetary heat radiation is  described by an effective temperature $T_{E}$ such that the total re-radiated power, $P_O$, is given by the well-known Stefan-Boltzmann law,

\begin{equation}
P_O=K_{S} T^4_{E}\, ,
\end{equation}

\noindent where the constant $K_{S}$ is given by the product of the surface area of the planet, $A$, the emissivity, $\epsilon$, and Stefan-Boltzmann's constant, $\sigma$. 
Note that $P_O$ is monotonously increasing as a function of the effective temperature with an everywhere positive second order derivative.

The expression for the total received power, $P_I$, involves the planetary "whiteness" i.e. albedo, and is assumed to depend on the effective temperature $T_E$. 
We have,

\begin{equation}
P_I=(1-\alpha) P_S\, ,
\end{equation}

\noindent where $P_S$ is the incoming power, e.g. stellar or solar radiation, and $\alpha$ the effective albedo. The latter is defined as the fraction of the received power which is re-radiated without prior absorption. 
We shall assume $\alpha \in ]0,1[$. 
Many simple functional forms of the albedo with temperature, e.g. being a constant $\alpha(T_E)=\alpha_0$, or having an everywhere negative curvature leaves only one solution for the requirement of heat balance, i.e.

\begin{equation}
P_I=P_O\, .
\end{equation}

\noindent More complicated forms for the albedo as a function of temperature will have one or a higher odd number of equilibrium solutions -- an even number of equilibrium points requires touching but non-crossing lines. In the general case, we have to solve the equation

\begin{equation}
(1-\alpha(T_E)) P_S=K_{S} T^4_{E}\, .
\end{equation}

\noindent Below, we shall study the case of three equilibrium points which can arise when the albedo function is loosely speaking an elongated Z-shape consisting of three straight pieces: At low temperature a horizontal line corresponding to the single phase of ice, at high temperature a horizontal line for the single phase of water, and finally an inclined line at intermediate temperatures representing coexisting phases, see Figure 1A. Sometimes an S-shaped function is used. For simplicity, we can construct both scenarios using a sigmoid-like function $f_i$, where 

\begin{equation}
f_0(x) = (1+e^{-x})^{-1}\, ,
\end{equation}
\noindent and
\begin{equation}
f_1(x) = \left\{ \begin{array}{lll}
0 & \mbox{if $x<-1/2$} \\
x+1/2 & \mbox{if $-1/2\leq x \leq 1/2$}\\
1 & \mbox{otherwise}
\end{array}
\right.
\end{equation}

\noindent The "albedo curve" is then chosen to be of the form,

\begin{equation}
1-\alpha(T_E)=\alpha_0+  (\alpha_{\infty}-\alpha_0)
f_i((T_E-T_0)/w)\, .
\end{equation}

\noindent For $i=0$ the standard S-shaped logistic function is recovered; for $i=1$ one obtains the elongated Z-shape. The constant $w$ is proportional to the inverse of the slope of the albedo curve at $T=T_0$. 
Any other sigmoid--like function, i.e. a monotonic function that interpolates between 0 and 1, with slope equal to one at $x=0$, could have been used with the same qualitative, though not quantitative, results.
The motivation for introducing this functional form for the albedo curve is the well-known scenario used in the snowball Earth hypothesis, namely, that the albedo of water can change drastically with its thermodynamic phase and hence can change continuously when the effective temperature allows for the coexistence of phases. Typically, this means an effective temperature which is not too far from the triple point ($T_0=273.16$K). 

Let us assume that we have the situation with three crossings, i.e. three equilibrium temperatures, as depicted in Figure 1A. The three points are denoted the S, the H, and the D points from the associations snowball planet, habitable planet, and desert planet. The corresponding effective temperatures are denoted $\{ T_S, T_H, T_D\}$. Both of the two dominantly single phase points, S and D, are in a stable equilibrium. The reason is that the Taylor expansion of $ P=P_I - P_O $ at these points 
have a negative first order derivative,

\begin{equation}
\partial_T P = \frac{\partial P(T^*)}{\partial T}  < 0 \,\,\,{\rm , where}\,\,\ T^* \in \{T_S, T_D\}
\, .
\end{equation}

\noindent From this, we observe that with a small positive temperature change, $\Delta T$,

\begin{equation}
P(T^* + \Delta T) \simeq \partial_T P \Delta T < 0\, ,
\end{equation}

\noindent therefore the planet is cooled, and hence the system is driven towards the stable equilibrium point, $T^*$.
This is not the case for the habitable point, H. At this multiphase point the first order Taylor coefficient $\partial_T P$ is positive and any deviation from equilibrium will be amplified over time. Hence, this is an {\it unstable} equilibrium solution. Can this unstable point exhibit relatively long-term stability?

Let us consider the axial inclination of a planet which has icecaps at the two poles (in our model, we assume that the icecaps are of the same size, see Figure 2B). This means that the incoming heat varies with the seasons of the year, i.e. it depends on how the planetary rotation axis is oriented relative the Sun or star. If we only consider the first order expansion of $P$, the integral of the incoming heat will be linear and the result will again be instability. It is necessary to consider the second and third order Taylor coefficients,

\begin{equation}
P(T + \Delta T) = P(T)+\partial_T P \Delta T  + \frac{1}{2} \partial_T^2 P(\Delta T)^2  + \frac{1}{6} \partial_T^3 P (\Delta T)^3+ \ldots
\end{equation}

\noindent The second order coefficient can help to form stability half of the year but causes a similar instability the other half of the year. 
What about the third order coefficient? Akin to the first- and second-order terms, neither the third- nor the higher-order terms are sufficient to create stability. Therefore, a time-dynamic effect is needed.

\begin{figure}[h]\centering
\includegraphics[width=7.0cm]{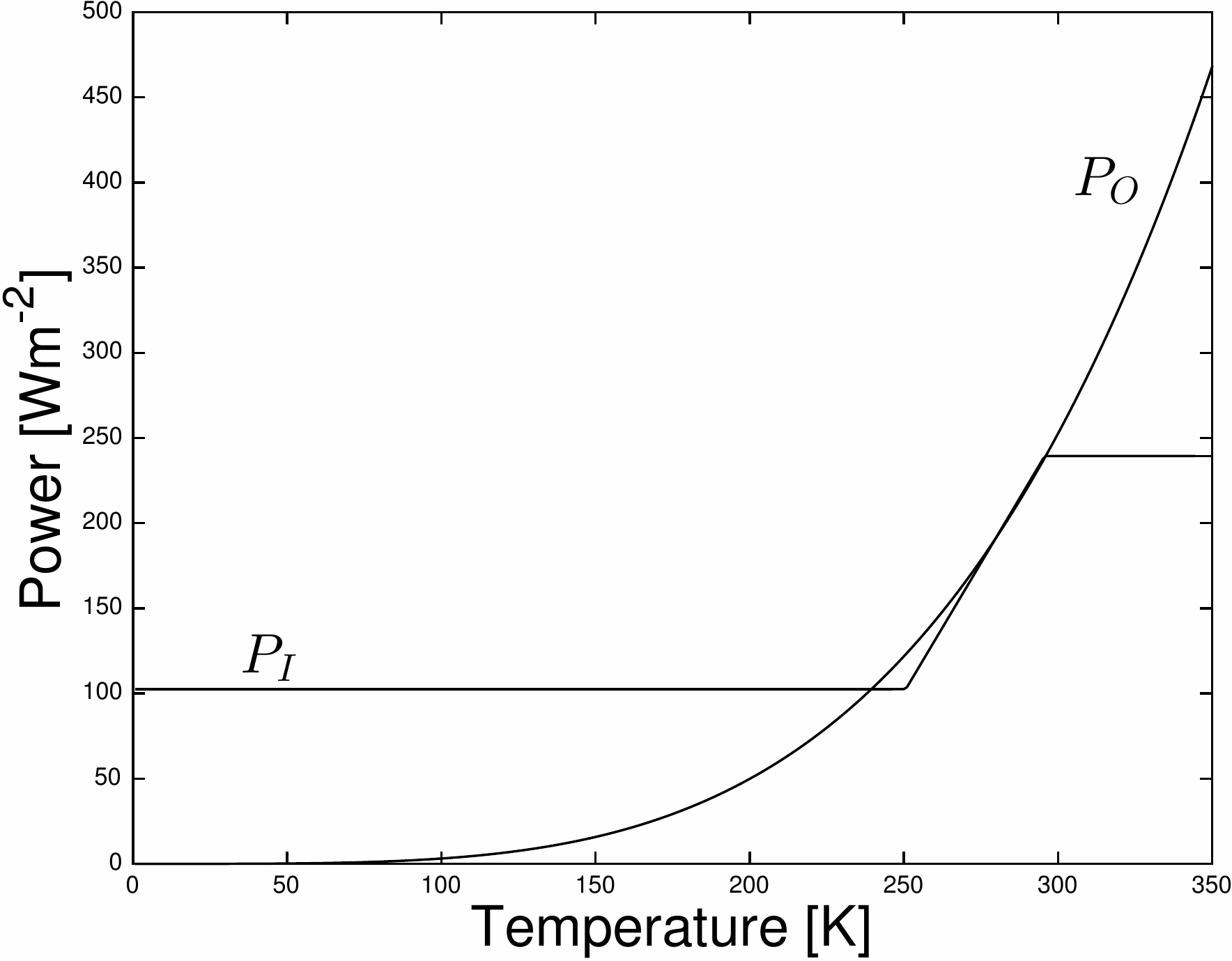}\includegraphics[width=7.0cm]{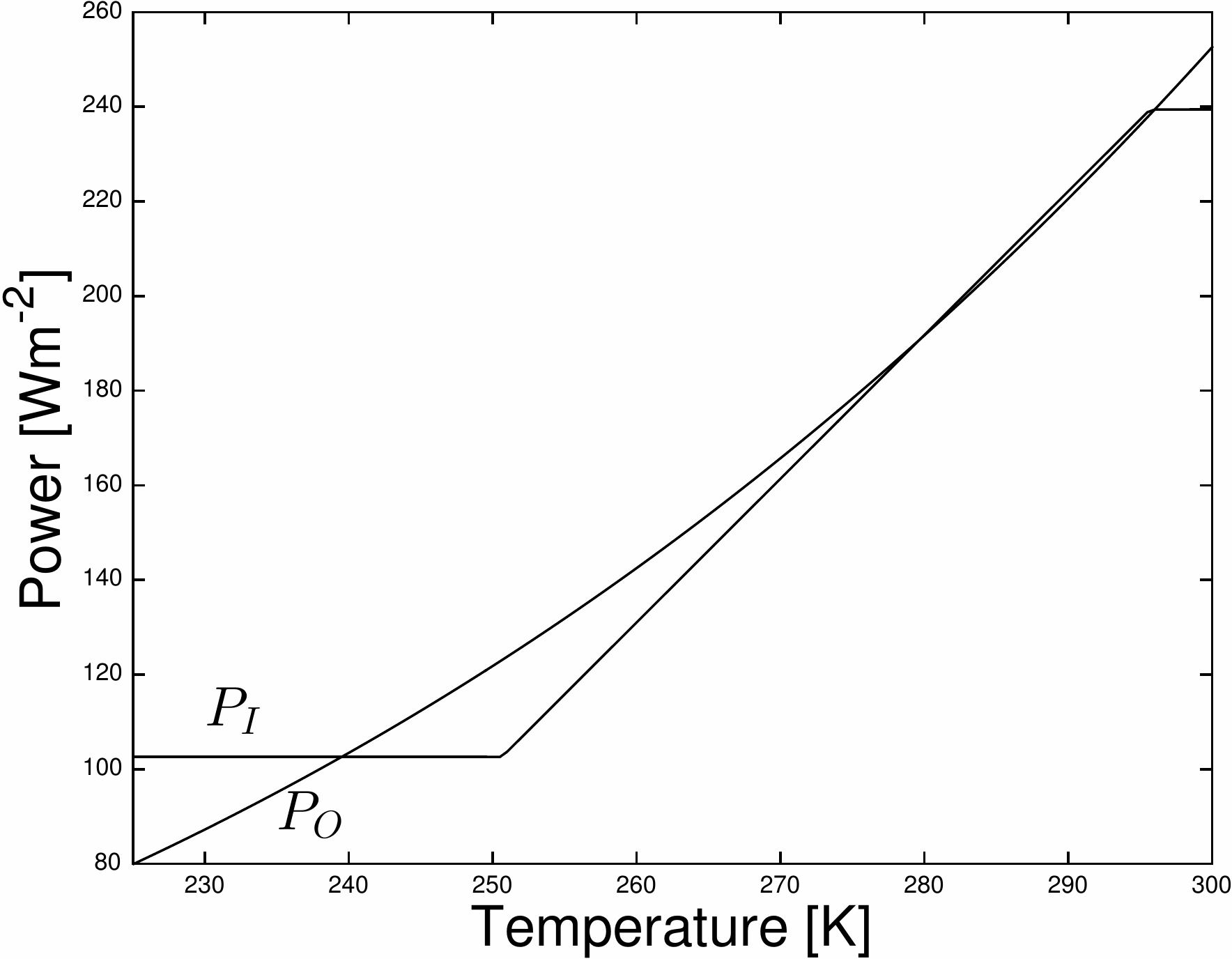}
\caption{Energy balance of the Earth: (A) received power $P_I$, here given per area, and radiated power $P_O$, also per area, as a function of effective temperature. $P_I$ is proportional to 1 minus the albedo, which at low and at high temperature corresponds respectively to complete ice cover, and no ice cover; (B) zoom of the energy balance, the three equilibrium temperatures are $\sim$  240K, 280K and 296K. An effective greenhouse factor of 0.55 was used in the calculations (see \cite{kaper2013}) and $w=10$ was used for the sigmoid function. The albedo is taken to be in the range,
$\alpha\in \left[ 0.3, 0.7\right]$.}
\label{xyz1}
\end{figure}

One time-dynamic effect is the presence of seasonal albedo variations.
Other factors which can contribute to periodic changes in the heat balance such as eccentricity of orbits are ignored and the planet is assumed perfectly spherical for simplicity. The received radiation at an area element ${\rm d}A$ is:

\begin{equation}
{\rm d} P_I (\theta ,\phi ,\Psi ,\Omega) = (1-\alpha(\theta )) \, \, H( \vec{R}(\theta ,\phi ,\Psi, \Omega ) \cdot \vec{S})\, S_0 \, \, {\rm d}A \, , 
\end{equation}

\noindent where $\vec{R}$ is the normal vector to the area element, $\vec{S}$ is the direction of the Sun or star, and $S_0$ the solar constant. $H$ is the Heaviside step function. The variable $\theta$ describes the planetary latitude while $\phi$ describes the longitude. The albedo is assumed to be rotationally symmetric and hence only depends on $\theta$. The tilt of the planetary axis (the obliquity) is denoted $\Psi$ and the angle $\Omega$ tracks the phase of the planetary position in its orbit. The average value of the received radiation can now be found by integrating over $\theta,\phi$ and averaging over $\Omega$ (a year), as $\Psi$ is perceived to be constant. For the Earth the current obliquity is $\Psi=23.4^{\circ}$. 

The seasonal variation of the absorbed power is found by avoiding to average over $\Omega$, see Figure 2A. This figure shows the two extrema, equinox versus summer/winter solstice. For ice caps of size $20^\circ$, the difference in fractional area, and therefore total power, is seen to be about $2\%$. Utilizing this observation, one can estimate the time constant related to seasonal changes (it is not simply one year!), and compare it to the time constant related to the instability of the habitable point: 
The effective global heat capacity of the Earth has been estimated in \cite{schwartz2007,schwartz2012} to be $C=21.8 \pm 2.1$ W yr m$^{-2}$ K$^{-1}$. The derivate of the total power with respect to temperature, $\lambda_j= (dP/dT)_j$, determines the time constant at an equilibrium point, as

\begin{equation}
\tau_j = C/\lambda_j\, .
\end{equation}

\noindent Using Figure 1, the three equilibrium points therefore have characteristic times constants of $\tau_S=12 (\pm 1), \tau_D=7 (\pm 0.6), \tau_H= -84 (\pm 8)$  yr. The first two of these are positive, while the one for the habitable point is negative (since it is an unstable point). However, the unstable point $H$ is also subject to seasonal variations. On the short time-scale, the seasonal variations correspond to a time constant $\tau_\Psi$. This time constant $\tau_\Psi$ is determined by the condition, that $1-2\% = 0.98 = \exp\left\{-(1/4\,  {\rm yr})/\tau_\Psi\right\}$, i.e. $\tau_\Psi \simeq 12$ yr. For the seasons to counter the habitable-point instability, we need approximately $|\tau_H| > 2\tau_\Psi$. As 84 ($\pm 8$) years is greater than $2\times12$ years, this is fulfilled. However, if we imagine that the seasonal effect was only three times smaller, then we would get $2\times\tau_\Psi= 75$ yr, and we could quickly be in trouble. In other words, the amount of axial inclination, the size of the ice caps and the orbital period are all crucial in having a stable climate.

\begin{figure}[h]\centering
\includegraphics[width=7.9cm]{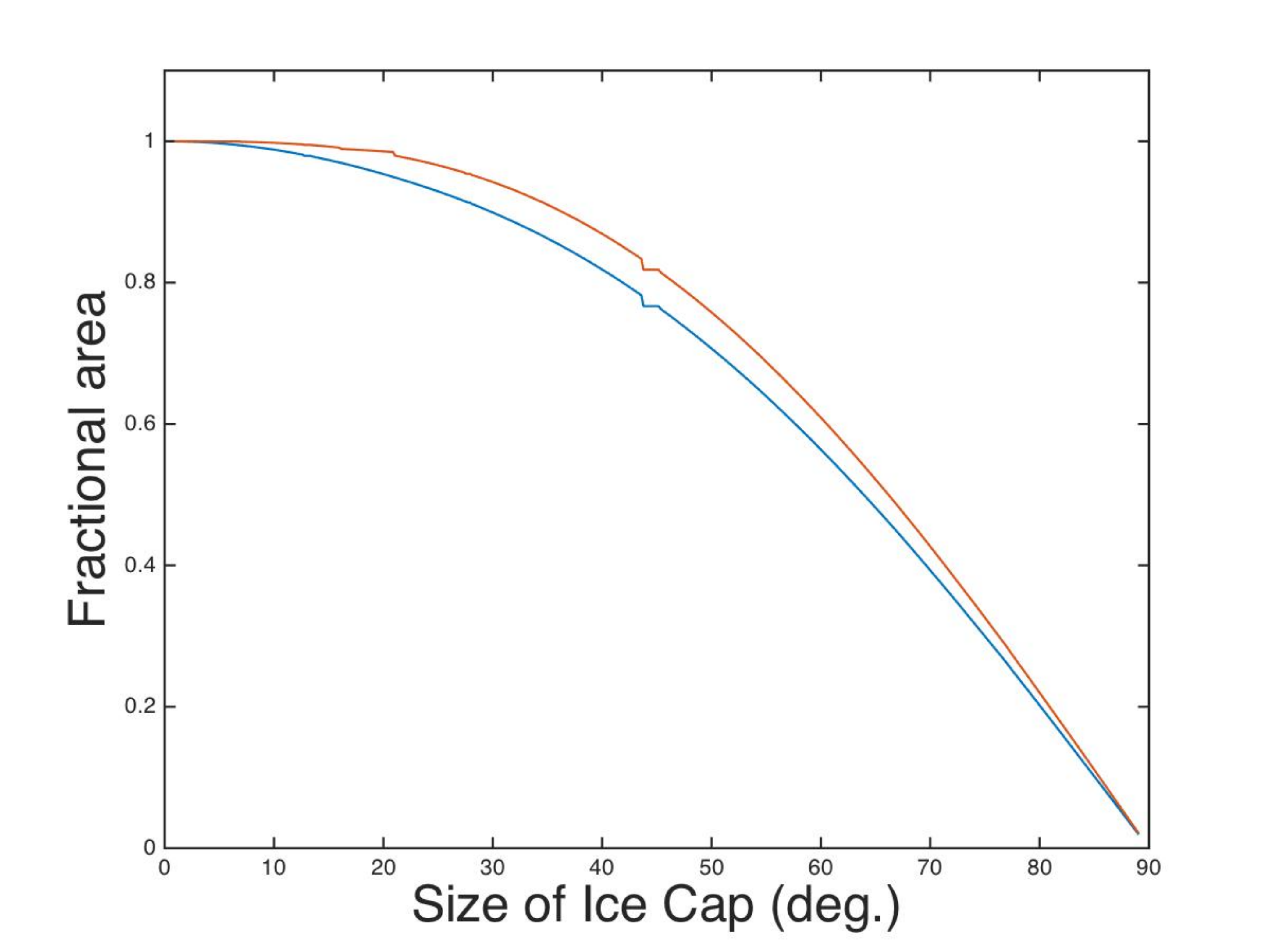}\includegraphics[width=4.9cm]{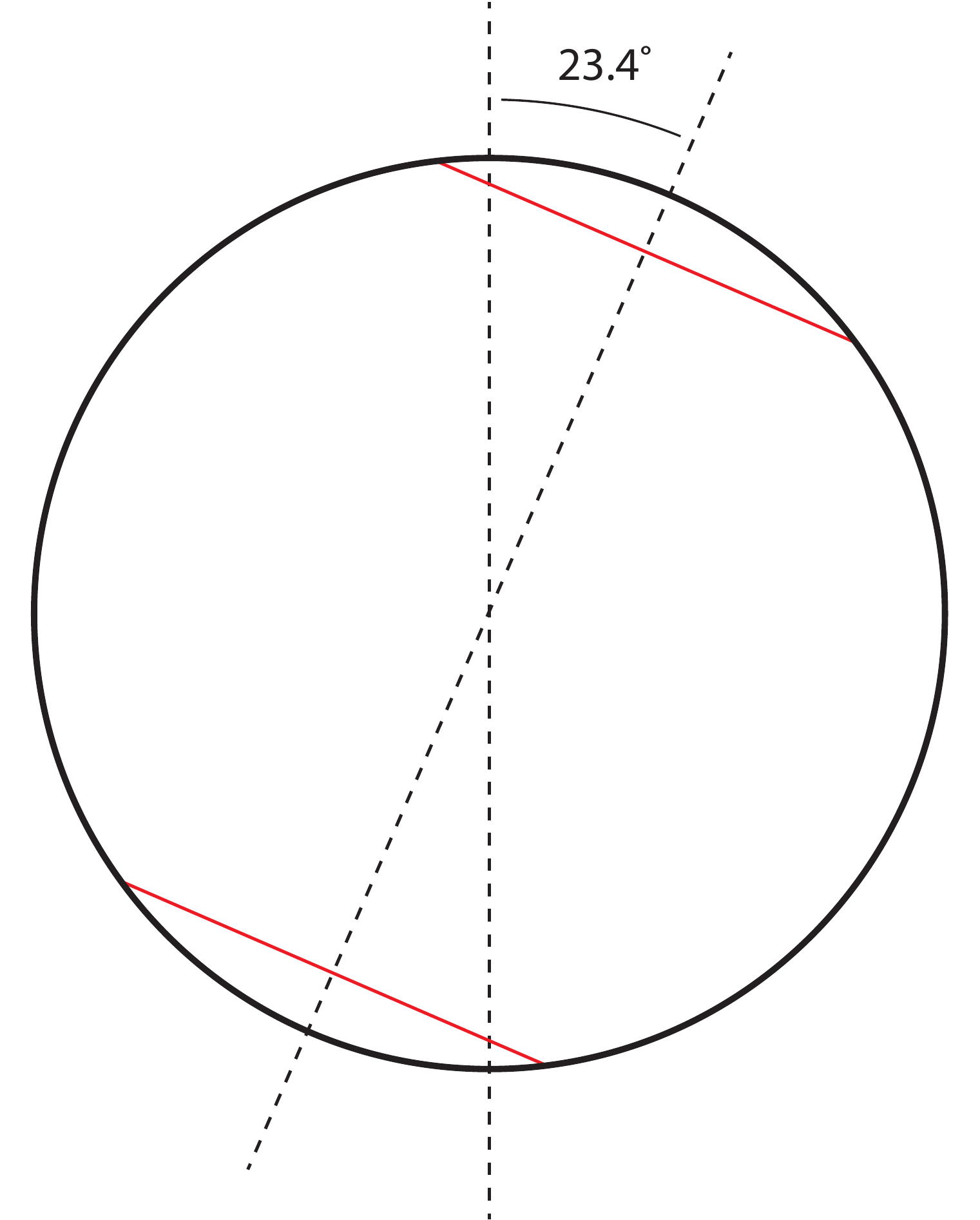}
\caption{(A) Fraction of the projected surface area with minimal albedo value ($\alpha=0.3$) as a function of the size of the two ice caps given in degrees -- at $90^\circ$ the planet is completely covered in ice. The upper curve (brown) is calculated at equinox, the lower (blue) curve at summer/winter solstice. The small dent at 45$^\circ$ is a numerical inaccuracy. (B) Our idealized Earth with an inclination of $23.4^\circ$. The ice caps are here illustrated with a size of $30^\circ$, the present size is about 20$^\circ$.}
\label{xyz2}
\end{figure}

\section{Discussion}

The suggested generic assignment of the habitable point deviate from previous assignments, where it has been suggested that the point which we call D is the habitable point \cite{kaper2013}. The rationale behind our chosen assignment can best be seen if the albedo function is taken in its most simple form, as in Figure 1A, corresponding to ice phase, co-existing phases and hot phase respectively.

The habitable point H is an unstable point in the static situation. However, because of the inclination of the axis of the Earth, it is a strongly driven system which leads to the seasonal changes of the climate: The heat capacity of the Earth leads to a delay in establishing equilibrium and hence a corresponding non-linearity. Indeed, strongly driven systems with forced oscillations can have stable equilibria \cite{zakrzhevsky2011,zevin1995}. Other papers have discussed the possibility of attaining stability by perturbations \cite{ott1990,shinbrot1993}.

Previously, there has been a discussion about whether eccentricity and planetary precession would lead to a planet temporary leaving the habitable zone, or otherwise having too extreme seasonal changes to support life as we know it from Earth \cite{williams1997,ashton2011,dressing2010}. These are valid concerns to be considered, yet at the same time one needs to incorporate the findings of the present work, namely that we shall treasure our seasons as it is their very existence that seem to stabilize the habitable climate. 

Our suggestion can also be important in the search for habitable exoplanets as many planets do not have a large angle of inclination. This will limit the number of observed exoplanets that will be candidates for habitation. However, we believe, that when searching long enough planets in the right belt with sufficiently inclined axis will be found to display a similar intermediate stability.

Finally, the stability analysis of equilibrium points has implications for the on-going climate debate.  It becomes evident that one cannot in all cases obtain a new and modified climate by smoothly adjusting the boundary conditions, as for example the amount of Greenhouse effect. The reason is that if the climate is modified beyond a certain point, a far instability will arrive and the planet is forced into either the desert planet point or into the snowball point, neither of which will be a gentle climate for the human species. 

\end{document}